\documentclass[prl,twocolumn,showpacs,amsmath,amssymb]{revtex4}
\usepackage{dcolumn}% Align table columns on decimal point
\usepackage[dvipdf]{graphicx}% Include figure files
\usepackage[dvips]{color}
\DeclareGraphicsExtensions{.jpg,.pdf,.mps,.png,.eps,.ps,.EPS}
\begin{document}
\def\be{\begin{equation}}
\def\ee{\end{equation}}
\def\bc{\begin{center}}       
\def\ec{\end{center}}
\def\bea{\begin{eqnarray}} 
\def\eea{\end{eqnarray}}
\newcommand{\avg}[1]{\langle{#1}\rangle}
\newcommand{\Avg}[1]{\left\langle{#1}\right\rangle}

\title{Dynamics of condensation in growing complex   networks}

\author{ Luca Ferretti$^1$ and Ginestra Bianconi$^2$, }
\affiliation{$^1$SISSA, via Beirut 4, I-34014 Trieste, Italy\\
$^2$The Abdus Salam International Center for Theoretical Physics, Strada Costiera 11, 34014 Trieste, Italy\\
}

\begin{abstract}
A condensation transition was predicted for growing   technological
networks evolving 
by  preferential attachment and   competing quality of their nodes, as described by the fitness model.
 When this
condensation occurs a node acquires a finite fraction of all the links
of the network. Earlier studies based on steady state degree distribution and on the mapping
to Bose-Einstein condensation,  were able to identify  the critical point.
Here we characterize the dynamics of condensation and  we present evidence that  below the condensation temperature there is a
 slow down of the dynamics and that a single  node ({\it  not} necessarily the
 best node in the network) emerges as the  winner for very long times.
The characteristic time $t^*$ at which this phenomenon occurs diverges
both at the critical point and at $T\rightarrow 0$ when new links are
attached deterministically to the highest quality node of the network. 
\end{abstract}
\maketitle

Condensation phenomena
\cite{Krapivsky,Bose,Rodgers,redner,Doro1,Doro2,Chayes} in complex networks \cite{Review,Doro_book,Vespignani,Doro_review} are  
structural phase transitions in which   a node 
grabs a finite fraction of all  the links of the network.
This phenomenon is of particular
interest in the case of technological networks.
The maps of the Internet  at the Autonomous System Level
show that the fraction of nodes connected to   the most connected node
is increasing in time reaching a share  the order of 
$10\%$ in recent maps \cite{nota}. Also in the World-Wide Web the share of webpages linked to Google webpages is of the order of $1\%$ a large number if one takes into account the size of the World-Wide-Web.
Are the Internet and the World-Wide-Web close to a condensation transition?
Which are the dynamical signatures of a
 condensation? 
Which are the consequences of a condensation of technological networks?
The problem might have relevant implications for the monitoring of
technological networks and there might be important difference between
the statistics of lead change below and above \cite{record} the condensation transition.In the following we are studying these problems focusing on the dynamics of the fitness model \cite{fitness,Bose} in which
nodes acquire links in proportion to their connectivity and in proportion of   their fitness  which indicates  the quality of the node. 
This model,   has been  considered a good stylized model for the
Internet \cite{Vespignani} and is a good stylized model describing the
emergence of high quality search engines in the   World-Wide-Web and
show a condensation phase transition depending on the parameters of
the model.
The study of the dynamical properties of this model will be able give
some estimates of the characteristic  time scale at which a condensate node
might  emerge in a condensation scenario and will allow us to   evaluate,
once  a condensate node is formed,  the 
 probability that a new node would overcome the condensate at later times.   

Questions concerning the dynamics of condensation in within the
fitness  model, might also  shed some light on the
relation between  the condensation phenomena occurring in this model an
the other off-equilibrium condensation phase transitions  occurring in  complex systems ( traffic jam,wealth distribution, urn and network models) 
\cite{Burda1,Evans1,Evans2,Godreche1,Godreche2,Mezard,Noh1,Noh2,Burda2}.
Off-equilibrium phenomena have attracted large interest in the last ten
years. Mainly  the attention has been addressed to  the  characterization
of the steady state of these out-of equilibrium systems, above the condensation transition \cite{Burda1,Evans1,Bose,Rodgers,Chayes}. On the contrary the dynamics by which the condensate emerges \cite{Evans1,Godreche1,Godreche2,Noh2} has been studied only in the case of urn models and models which can be reduced to the Zero Range Process \cite{Evans2}. 
However the models where the condensation occurs as a structural phase transition of a network are not in general reducible to the
Zero-Range-Process.

In this paper we study the dynamics of condensation in the fitness
model. We apply the rank statistics for the description of the
dynamics of condensation and we will show that below the condensation
phase transition a condensate
node emerges only after a characteristic time $t^*$. This node is then one of the nodes with highest fitness. The probability that later-comers high fitness nodes will overcome the condensate
is decaying with time and there is a slow down of the dynamics.
At the condensation transition the characteristic  time $t^*$ diverges
and the condensation is only marginal, i.e. we have a sequence of
highly connected high-fitness nodes each one overcoming the other
 each of then grabbing a very small fraction of the total links
of the network.

{\it The fitness model and the condensation phase transition-}
The fitness model \cite{fitness} is a growing network model.
We start from a finite connected network of $N_0$ nodes.
At each time $t_i$ a new node $i$ and $m<N_0$ new links are added to the
network. To the node $i$ it is  assigned a  quenched variable,
$\varepsilon_i$ ('energy' of the node ) drawn from a $g(\varepsilon)$ distribution.
The variable $\varepsilon_i$ indicates
the intrinsic quality  of the node
(lower 'energy' better quality). Each of the $m$ new links that are
added to the network at time $t_i$,  
 connects the new $i$ node to a node $j$ chosen with probability  
\be
\Pi_j(t)=\frac{\eta_j k_j(t)}{\sum_{\ell} \eta_{\ell} k_{\ell}(t)}.
\label{pi}
\ee
where we have introduced  the fitness $\eta_j$ associated to  node $j$  as
\be
\eta_j=e^{-\beta \varepsilon_j}.
\ee
The parameter $\beta=1/T$ tunes the relevance of 
the quality of the nodes in the choice of the target node.
In the limit case $T\rightarrow \infty$ the dynamics of the attachment
of new links is only dependent on the connectivity of the nodes and we
recover the Barab\'asi-Albert  model \cite{BA}.
For $T\ll 1$ the intrinsic quality of the nodes highly affects the dynamic of
the system.

\begin{figure}
\includegraphics[width=80mm, height=45mm]{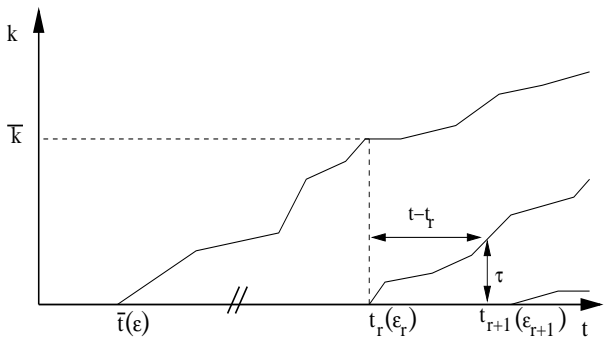}
\caption{We plot the connectivities of records as a function of time. The time $t_r$ indicates the time at which a record node with $\varepsilon_r$ arrive in the network.
The effective model only focuses on the competition for links of the most connected nodes with the node of highest fitness.
\label{effective}}
\end{figure}

For the  $g(\varepsilon)$ distribution such that $g(\varepsilon)\rightarrow 0$
as $\varepsilon \rightarrow 0$  a phase transition can occur at a
critical temperature $T_c$. Above this critical temperature every node has an infinitesimal fraction of all the links, below the critical temperature one node grabs a finite fraction of all the links.
These results can be obtained by consideration of the characteristics
of the steady state of the model above the phase transition. In fact the model can be solved in a mean-field approximation by making a
self-consistent assumption. In particular in \cite{fitness} it is assumed that 
$Z_t=\sum_j \eta_j k_j(t) \rightarrow \avg{Z_t}\rightarrow me^{-\beta \mu}t+{\cal O}(t^0)$.
With this assumption it can be shown  that, at time $t$, the 
average connectivity of node $i$ arrived in the network at
time $t_i$  grows as a power-law 
$\avg{k_i(t)}=m\left({t}/{t_i}\right)^{f(\varepsilon_i)}$
with $f(\varepsilon)=e^{-\beta(\varepsilon-\mu)}$ and with  the constant $\mu$ determined  by the self-consistent equation

\be
1=I(\beta,\mu)=\int d\varepsilon
g(\varepsilon)\frac{1}{e^{\beta(\varepsilon-\mu)}-1}.
\label{selfc}
\ee
In \cite{Bose} it was shown  that  the  self-consistent equation $(\ref{selfc})$ is equivalent to the equation for the
conservation of the total number of links present in the network
\be2mt=mt+mtI(\beta,\mu).\ee
If Eq. $(\ref{selfc})$ has a solution then the degree distribution of
nodes of fitness $\eta=e^{-\beta\varepsilon}$ is given by
$P_{\varepsilon}(k)\simeq k^{-1/f(\varepsilon)-1}$ and $P(k)=\int d\varepsilon
g(\varepsilon) P_{\varepsilon}(k)$ is dominated by the term
$P_0(\varepsilon)=k^{-\gamma}$ with $\gamma=e^{-\beta\mu}+1$ but can
have logarithmic corrections \cite{fitness}.
Nevertheless, if $I(\beta,0)<1$ the self-consistent equation  $(\ref{selfc})$ cannot be
solved. In this case the self-consistent approach
fails and we have the condensation phase transition.
A necessary condition for condensation,( i.e for $I(\beta,0)<1$)  is
that the distribution $g(\varepsilon)\rightarrow 0$
for $\varepsilon \rightarrow 0$.
For this type of $g(\varepsilon)$ distributions there can be  a critical temperature
$T_c=1/\beta_c$ such that $I(\beta,0)<1 \ \  \forall \ \ \beta>\beta_c$.
This phase transition is formally equivalent to the Bose-Einstein
condensation for non-interacting Bose gases in dimensions $d>2$.
Using the similarity to the Bose-Einstein condensation in Bose gases,
in Ref. \cite{Bose} it was assumed  that below the condensation temperature  the equation for conservation
of the links can be written as
\be
2mt=mt+mtI(\beta,0)+n_0(\beta)mt
\label{n00}
\ee
where $n_0(\beta)$ indicates the number of links attached to the most
connected node.
Consistently with $(\ref{n00})$, simulation results reported in 
 \cite{Bose} show that indeed the fitness model undergoes a
 condensation phase transition at $T=T_c$. 
An example of $g(\varepsilon)$ distributions for which there is a
condensation is 
\be
g(\varepsilon)=(\theta+1)\varepsilon^{\theta} 
\label{distr}
\ee
and $\varepsilon\in [0,1]$. Here and in the following we will always
consider $g(\varepsilon)$ distributed according to $(\ref{distr})$.
Assuming $(\ref{n00})$ and a distribution of the type $(\ref{distr})$
the fraction of condensed links, similarly to the corresponding result
for the Bose gas, will take the form 
\be
n_0(\beta)=1-\left(\frac{T}{T_c}\right)^{\theta+1}.
\label{sn}
\ee
\\
{\it Dynamics of the fitness model-}
 The necessary condition for condensation
$g(\varepsilon)\rightarrow 0$ as $\varepsilon \rightarrow 0$ is a clear
indication that dynamical effects not captured by the stationary state approaches
\cite{Bose,Rodgers,Chayes} will be very important for the emergence of
a long lasting condensate node in the network. Moreover all considerations
regarding the dynamical aspects of the condensation phenomena on
the fitness network cannot take advantage to the similarity  with the
condensation phenomena in the non-interacting Bose gas.
In fact, in  a Bose gas, the equivalent of  $n_0(\beta)$ is the
occupation of the state at energy $\varepsilon=0$. On the contrary, in the fitness networks, $n_0(\beta)$
cannot be the connectivity of the fittest node ($\varepsilon=0$).
Indeed, in  the fitness model, the fittest node appears in the network only in the infinite
time limit because the probability $g(\varepsilon)$  of having a node of quality
$\varepsilon$, goes to zero as $\varepsilon\rightarrow 0$.
Therefore the question is: is the condensate node the best node in the network? 
Once the condensate has appeared in the network is there a possibility
for later-comers high fitness nodes to take over the condensate?
To narrow down this general question we define a specific problem we
want to address.
First of all, let  us introduce the notion of a
  record. A record  $\varepsilon_R(t)$ at time $t$ is the node (or
equivalently the value of its $\varepsilon$) with minimal
$\varepsilon$ (maximal fitness) present in the network.
In the fitness model below the condensation transition, all records
are good candidates as condensate nodes. 
Suppose that   a record $\varepsilon$, arrived in the network at time
$\bar{t}$, is a condensate with
$\bar{k}$ links at the time $t_{r}$  where a the $r$-th record occurs. We
want to know the probability that  the $r$-th record of fitness
$\eta_r=e^{-\beta \varepsilon_r}$    overcome the
condensate node before the subsequent record $\varepsilon_{r+1}$ enters  the
network (See Fig. $\ref{effective}$ for a pictorial representation of
our problem).
In order to answer this question we have to use some results related
to record statistics.
\\
{\it Records-}
The typical value of the record at time $t$ is given by
\be
\int_0^{\varepsilon_R} g(x) dx=\frac{1}{t}, 
\label{records}
\ee indicating that the average number of nodes   with
$\varepsilon>\varepsilon_R(T)$  in a network of $t$ nodes is less than
one. Using $(\ref{records})$ for the distribution $(\ref{distr})$ we
obtain 
\be 
\varepsilon_R(t)\simeq t^{-\frac{1}{\theta+1}}.
\label{epsilonr}
\ee
The statistics of records is a field on its own \cite{Records}, here
we cite only another result which will be used in the following: given
the $r$-th  record at time $t_r$ the
probability to have the following record at  time $t_{r+1}>t$   in the
approximation of a continuous time is given by \cite{Records}
\bea
P_R(t_{r+1}>t,t_r)=\frac{t_r}{t}
\label{rec}
\eea
where $t>t_r$.
\\
{\it The effective model-}
At low temperatures, below the condensation phase transition, the
linking probability $\Pi_j$ (Eq.$(\ref{pi})$) of the fitness model will be relevant only for nodes of
  very high fitness (the series of records) or for the condensate node. Therefore we consider
  the extremely simplified effective model of just  two  competing
  nodes (the condensate node and the record) disregarding all the
  other nodes present in the network. This simplification will provide
  us a scenario for condensation which we will subsequently compare
  with simulation results on the original fitness model.
At time $t_{r}$ there is a node (the condensate) arrived in the
network at time $\bar{t}$ with
fitness $\bar{\eta}=e^{-\beta \bar{\varepsilon}}$ and $\bar{k}$ links, and a record node with fitness
$\eta_r=e^{-\beta \varepsilon_r}$ which is just born and has degree
1. At each time we distribute one link to one of the two nodes with
probability given by expression $(\ref{pi})$.

The probability $p_t(\tau)$ that the node with fitness $\eta_r$ has
$\tau$ links at time $t$ satisfies in the continuous time limit
 \begin{equation}
\frac{\partial p_t(\tau)}{\partial t}= \frac{\eta_r
  (\tau-1)}{Z_{t,\tau -1}}p_t(\tau -1)-\frac{\eta_r \tau}{Z_{t,\tau}}
p_t(\tau)\label{eqp},
\end{equation}
where  the competition is included in the formula through
$Z_{t,\tau}=\bar{\eta}(\bar{k}+t-t_r-\tau)+\eta_r \tau$.
We are interested in the dynamics of  this  effective model  until
the node with fitness $\eta_r$ takes over, i.e. until
\be
\tau<\tau_C=(\bar{k}+t-t_r)/2.
\ee
 Equation
$(\ref{eqp})$ is non-trivial to study and some approximations are
necessary. Therefore we assume  that  $Z_{t,\tau}\simeq Z_{t,\tau-1}$,
(i.e. $\beta \bar{\varepsilon} \ll 1$) which, taking into account the scaling
of the records with the time of their appearance,
Eq. $(\ref{records})$ reduces to the condition 
\be
\bar{t}>T^{-(\theta+1)}.
\label{9}
\ee 
Furthermore we assume that 
 $C\simeq Z_{t,\tau}/(\bar{k}+t-t_r)$ can be approximated by a constant for
 $\tau<\tau_C$. Indeed for  $\tau<\tau_C$ we have that $C$ varies in a
 narrow interval
 $C\in(\bar{\eta},(\bar{\eta}+\eta_r)/2)$. In this approximation   we can rewrite  equation $(\ref{eqp})$ using the generating function $q(z,t)=\sum_{\tau=1}^\infty p(\tau,t) z^\tau$. The equation becomes 
\be
 C (\bar{k}+t-t_r)\frac{\partial q}{\partial t}=\eta_r\, z(z-1)\frac{\partial
   q}{\partial z}
\ee
which, if we impose the initial condition $q(z,{t}_r)=z$, has the solution 
\be
q(z,t)=\frac{z}{z\left(\frac{\bar{k}+t-t_r}{\bar{k}}\right)^{-\eta_r/C}+(1-z)}\left(\frac{\bar{k}+t-t_r}{\bar{k}}\right)^{-\eta_r/C},\nonumber\ee
and the expression for the probability $p_t(\tau)$ becomes
\begin{equation}
p_t(\tau)=\left(\frac{\bar{k}}{\bar{k}+t-t_r}\right)^{\eta_r/C}\left(1-\left(\frac{\bar{k}}{\bar{k}+t-t_r}\right)^{\eta_r/C}\right)^{\tau-1}\label{prob}.\nonumber 
\end{equation}
The probability that, at some time $t$, $\tau<\tau_C$, i.e. the
probability that the record has not become the most connected node  of the network   can be calculated:
\bea
P_1\left(\tau<\tau_C\right)&\simeq
&1-\exp{\left[-\frac{\bar{k}}{2}\left(\frac{\bar{k}}{\bar{k}+t-t_r}\right)^{(\eta_r-C)/C}\right]}\nonumber
\label{P1}
\eea
where in the last step we have assumed that $t_C-t_r \gg 1$. 
Expression $(\ref{P1})$ for $P_1(\tau<\tau_C)$ allow us to evaluate
the characteristic time $t_C$ at which the record $r$ becomes the most
connected node, i.e.
\be
t_C-t_r\simeq \bar{k}^{1+C/(\eta_r-C)}.
\ee
Therefore, by making use of  $(\ref{rec})$ we can evaluate  the probability $P^{\mbox{eff}}_C$ that the record $r$ has become the most connected node of the network before
the appearance of the record  $r+1$,
\bea
{P^{\mbox{eff}}_C}&=P_R(t_{r+1}>t_C,t_r)=&\simeq\frac{1}{\hat{n}
  \bar{k}^{C/(\eta_r-C)}}=\frac{1}{\hat{n} \bar{k}^{\alpha}}
\label{Pscale}
\eea
where $\hat{n}=\bar{k}/t_{r}$.
Taking into account  the interval of definition of  $C$, the scaling exponent $ \alpha=C/(\eta_r-C)\in[ 1/[\beta(\bar{\varepsilon}-\varepsilon_r)],
2/[\beta (\bar{\varepsilon}-\varepsilon_r)]]$. Therefore if
$\bar{\varepsilon}\gg\varepsilon_r$ then  
\be 
\alpha \simeq \frac{1}{\beta \bar{\varepsilon}}\simeq1
 T{\bar{t}}^{\frac{1}{\theta+1}}.
\ee
In other words as long as Eq. $(\ref{9})$ is satisfied,  the exponent $\alpha$ is greater than
one, (i.e. $\alpha>1$) and it increases with larger  $\bar{t}$ times
and for higher temperatures $T$.

\begin{figure}
\includegraphics[width=80mm, height=50mm]{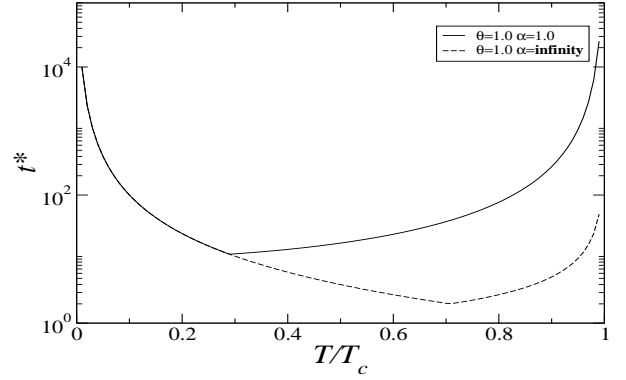}
\caption{Characteristic time $t^*$  calculated with the estimation
  $(\ref{est})$ (B=10) after which a long lasting
  condensate node emerges for the fitness model for  $\theta=1$ and
  $\alpha=1, \infty$
\label{tstarfig}}
\end{figure}

If we fix the condition $P_C^{\mbox{eff}}<1/B$
we find a characteristic time for the appearance of a long lasting
condensate node  in the
effective model is given by 
\be
 t^*=\max\left(T^{-(\theta+1)},\left(\frac{B}{\hat{n}}\right)^{1/{\alpha}}\frac{1}{\hat{n}}\right)
\label{ts1}
\ee
where $\hat{n}=\bar{k}/t_r$.
\begin{figure}
\includegraphics[width=80mm, height=50mm]{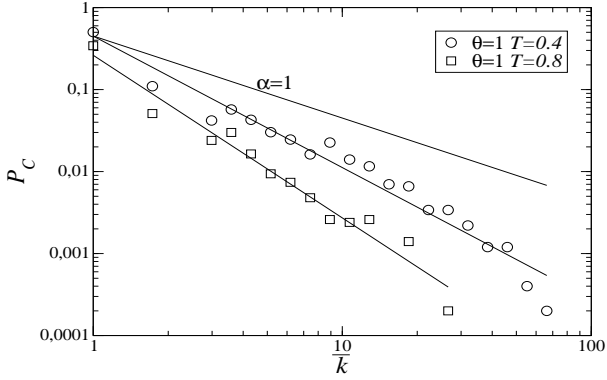}
\caption{Probability $P_C$ that a record overcomes the most
  connected node of the network in the fitness model averaged over
  $10^4$ runs for a time lapse of $t_{max}=10^5$.\label{P}}
\end{figure}
\\
{\it Dynamics of condensation-}
We expect that the effective model, describing the competition of only
two nodes, provides an upper bound for the probability $P_C(\bar{k})$
that in the fitness model a new
record overcomes the most connected node before the appearance of a
subsequent record.
We define  $P_C(\bar{k})$  the probability that a record, introduced in
a network with a condensate of connectivity $\bar{k}$, is able to overcome
the connectivity of the condensate node before the subsequent record. To numerically
evaluate $P_C(\bar{k})$ we consider all the records that follow the
emergence of a condensate and we evaluate the ratio between the number
of  positive events and the total number of records that appear in the network
when the condensate has ${\bar{k}}$ links.
The data reported in Figure ${\ref{P}}$
are taken   for $\theta=1$ at different temperatures $T$  below the phase
transition. The results shown are statistics taken  over
$10^4$ networks  for a time lapse $T^{-(\theta+1)}<t<10^5$.
Figure $\ref{P}$ shows clearly  that the probability ${P}_C(\bar{k})$   decays
as 
\be
P_C(\bar{k})\propto \bar{k}^{-\alpha}
\ee
 with $\alpha \ge 1$ 
and that the exponent $\alpha$ increases with $T$ as predicted by the
effective model.
Using the effective model estimate $(\ref{ts1})$ for  the characteristic time $t^*$ for having a long lived
winner node which condenses, and assuming $\hat{n}\simeq n_0(\beta)$
with $n_0(\beta)$ given by $(\ref{sn})$ we get 
\be
 t^*=\max\left\{T^{-(\theta+1)},B^{1/\alpha}\left[1-\left(\frac{T}{T_c}\right)^{\theta+1}\right]^{-(1+1/\alpha)}\right\}.
\label{est}
\ee
This characteristic time is depicted in Figure $\ref{tstarfig}$ for
different values of $\alpha$ and the results are consistent with a  
   divergence of $t^*$ at $T\rightarrow T_c$ and at
$T\rightarrow 0$.
 In order to evaluate  the time life of a condensate for  times
 $t>t^*$ we can use  the scaling of the probability  $P_C \sim n_0^{-1}
\bar{k}^{-\alpha}$ and we can  assume that for the condensate node
$\bar{k}\simeq n_0 t$. Moreover if the relevant competition is only
between the condensate node and the node with highest fitness,  the  probability to have the
same condensate after  $S$ records is given by
\bea
{P_W^{\mbox{eff}}(S)}&\simeq &\prod_{s={r}^{\star}+1}^{{r}^{\star}+S} \left(1-\frac{1}{n_0(n_0 t_s)^{\alpha}}\right).\nonumber
\eea
where $t_s$ are the times of record occurrence after the condensate has
become the most connected node in the network (i.e. for $s>{r}^{\star}$).
The average of this quantity over the distributions of the times of the
records $\prod_s
P_R({t}_s,t_{s-1})=\prod_s t_{s-1}/t_s^2$ gives
\be
\Avg{P_W^{\mbox{eff}}}(S)\simeq
\exp\left\{{-\left[1-\frac{1}{(1+\alpha)^S}\right]\frac{1+\alpha}{\alpha
      n_0(n_0t_{r^{\star}})^{\alpha}}}\right\}.
\ee 
Therefore for long times, we expect $\avg{ P_W^{\mbox{eff}}}(S)\rightarrow
\mbox{const}$ as $S\rightarrow \infty$.
Consequently we can conclude that for times $t>t^*$ the condensate
node is typically  not the best node
of the network nevertheless  it dominates the network for very long times.
Only for  $T\rightarrow T_c$  and $T\rightarrow 0$ the time needed to
havelong lasting  condensate  diverges and in the network there is a fair
competition and a  succession of high fitness nodes on which
condensation occurs.

{\it Conclusions-}
In conclusion we have studied the dynamics of condensation in growing
network models  within the  fitness model. We show that below the
condensed phase transition there is a characteristic time $t^*$. For times $t<t^*$ the
network dynamics is dominated by a succession of high fitness nodes
with a finite fraction of the links, above the characteristic time
$t^*$ a long-lasting winner takes over acquiring a finite fraction of all the
links and slowing down the dynamics. The characteristic time $t^*$ is
diverging at the critical point of the condensation phase transition
and in the limit $T\rightarrow 0$. These results may have strong
implications for the monitoring of technological networks in which we
expect a fair competition lasting for very long times only if close to
the condensation transition.

We acknowledge fruitful discussions  with Claude Godr\'eche and Jean
Marc Luck and Marcello Mamino.

\end{document}